\documentclass[MSNbibl,nameyear,dvips]{arxstspdf}
\usepackage{flushend}
\usepackage{stfloats}

\volume{27}
\issue{3}
\pubyear{2012}
\firstpage{348}
\lastpage{349}
\doi{10.1214/12-STS381REJ} 
\referstodoi{10.1214/11-STS381}

\begin{document}
\begin{frontmatter}
\vspace*{6pt}
\title{Rejoinder}
\runtitle{Rejoinder}

\begin{aug}
\author[a]{\fnms{William} \snm{DuMouchel}\corref{}\ead[label=e1]{bill.dumouchel@oracle.com}}
\runauthor{W. DuMouchel}

\affiliation{Oracle Health Sciences}

\address[a]{William DuMouchel is Chief Statistical Scientist, Oracle Health Sciences, Van De Graaff Drive, Burlington, Massachusetts 01803, USA
\printead{e1}.}

\end{aug}


\vspace*{-3pt}
\end{frontmatter}

Many thanks to all of the discussants for their observations and
suggestions and encouraging words. My response is quite short because I
agree with all of their comments.

Professor Evans points out the importance and potential difficulty of
selecting which set of issues or adverse event definitions to include in
the analysis. They need to be approximately exchangeable in terms of
prior belief as to their expected associations with Treatment and with
covariates. Prior belief refers to your expert opinion before looking at
the data to be analyzed---it would not be appropriate to just select
issues that are observed to have high odds ratios with treatment, for
example, which might tend to bias the estimation of variance components
in the model, making them appear too small. His suggestion of selecting
SMQs that are formed from terms in a local region of the MedDRA
hierarchy seems reasonable, or of using other knowledge or data that
seem to point to plausible exchangeability. I also agree with Professor
Evans on the need for more experience with how MBLR works with varying
numbers of issues ($K$) and numbers of covariates ($J$) and varying sample
sizes. My intuition is that the model should be robust to a wide range
of $K$ and sample sizes, but that it is probably not a good idea to allow
$J$ to be too large, on the rationale that collinearity often leads to
problems in any regression model having many covariates. Perhaps $J$
should be kept in the range of 1 to 5, and I suspect that it would be a
mistake, leading to spurious results, to think of MBLR as a data mining
technique, where you search for the best 5 covariates out of 50
available, for example. Better to just use a few prespecified covariates
that have some prior justification for potential effects or interactions
with treatment.

Professor Berry provides a nice overview of the inferential and medical
difficulties raised by drug safety issues. Many of these difficulties
are beyond the reach of any mere statistical model, of course. But in as
much as consideration of multiplicities due to analyses of related
medical events and to subgroup analyses contribute to these problems,
perhaps the MBLR model, or similar approaches using Bayesian shrinkage,
can help researchers see the forest through the trees.

Special thanks go to Brad McEvoy and Ram Tiwari for their collaborations
and discussions during the later development of MBLR, and now for their
acute comments here. First, they provide an insightful and enlightening
discussion of the relationship of the MBLR model to meta-analysis
modeling in general, assuming that the data at hand come from multiple
studies. My choice, made for simplicity, was to treat the Study ID as
just another categorical variable like patient age or sex. This should
eliminate certain biases like those due to Simpson's paradox, and does
allow the treatment effect to vary by Study, but does not treat Study as
an independent random effect that could potentially interact with all
other covariates. It also makes it difficult to build into the analysis
the fact that patients were independently randomized within each study,
which does not play much of a role in a Bayesian model, but is an
important aspect of the classical paradigm. McEvoy and Tiwari define an
extended meta-analytic MBLR model that allows for Study (trial) to be a
higher level of a hierarchical model and fit it to the data in the paper
using the Markov chain Monte Carlo (MCMC) approach. That model is much
more complex, since instead of just 4 prior standard deviations to
estimate, there are $(2K+4) = 24$ such hyperparameters, which would make
it impossible to fit using the simpler methodology of the present paper.
Much more research would be needed to assess the relative validity and
reliability of this more complicated methodology. An alternative
approach might be to fit the current MBLR model separately to the data
from each study, and then using a more traditional meta-analysis
approach as a post-processing step applied to the coefficients and
standard errors so obtained.

Drs. McEvoy and Tiwari also focus on the decision to borrow strength
across different issues (AEs), in that if your prior belief that the
issues are exchangeable is incorrect, then the shrinkage may not help
and could possibly lead to extraneous variability. They point out that
careful interpretation is especially needed when a rare but severe issue
is analyzed in conjunction with other more frequent but less important
or interesting issues. This is always a potential problem whenever a
more available variable is being treated as a possible marker for a less
available but more medically severe event or process.

Finally, these discussants used the MCMC method to re-estimate the MBLR
coefficient estimates from\ the example data and show that they are
remarkably similar to those based on my Laplace approximation method. We
have replicated this result using in-house computations, using the same
and similar models.

Once again, many thanks are due to all of the discussants for their
interesting and insightful and generous comments.

\end{document}